\documentclass[showpacs,pre,showkeys]{revtex4}
\usepackage{graphicx}

\usepackage[usenames]{color}
\usepackage{cancel}
\usepackage{hyperref}
\usepackage[normalem]{ulem}
\usepackage{amsfonts}
\usepackage{amsmath}
\usepackage{graphicx}
\usepackage{epstopdf}
\usepackage{float}
\usepackage{enumerate}

\begin{document}

\title{Widely distributed clusters of the constraint satisfaction problem model d-k-CSP}

\author{Wei Xu$^1$}
\author{Fuzhou Gong$^2$}
\author{Guangyan Zhou$^3$}
\email[]{zhouguangyan@btbu.edu.cn}
\affiliation{
$^{1}$School of Mathematics and Physics, University of Science and Technology Beijing, Beijing 100083, China\\
$^{2}$Institute of Applied Mathematics, Academy of Mathematics and System Science, Chinese Academy of Sciences, Beijing 100190, China\\
$^{3}$Department of Mathematics, Beijing Technology and Business University, Beijing 100048, China
}

\date{\today}

\begin{abstract}
Relation between problem hardness and solution space structure is an important research aspect. Model d-k-CSP generates very hard instances when $r=1$ and $r$ is near 1, where $r$ represents normalized constraint density. We find that when $r$ is below and close to 1, the solution space contains many widely distributed well-separated small cluster-regions (a cluster-region is a union of some clusters), which should the reason that the generated instances are hard to solve.
\end{abstract}
\maketitle

\section{Introduction}

The constraint satisfaction problem, or CSP, is an important topic in the interdisciplinary area of statistical physics, computer science and information theory. In the theoretical aspect, CSPs play a significant role in understanding the problem hardness as well as properties of disordered systems.  In practical use, CSPs are applied to many tasks, such as timetabling, hardware configuration, transportation scheduling, factory scheduling, floorplanning, error-correcting code, etc. A CSP formula contains variables and constraints, where the variables can be assigned a value from their respective domains, and the constraints contain a set of variables called the constraint scope that restricts their allowed joint values. One elementary question is to assign all variables such that all constraints are satisfied simultaneously; meanwhile, another elementary question is to determine whether a solution exists.

Popular CSPs, such as K-SAT and Coloring, have been intensely studied, and fruitful results have been obtained. Cheeseman {\it et al}, in \cite{chee}, found that many CSPs undergo a satisfiable-unsatisfiable transition: when the constraint density is low, almost every instance has a solution; but when the constraint density increases and exceeds an inherent point, the solutions suddenly disappear for almost every instance. Friedgut {\it et al} supported this opinion and proved that satisfiability of K-SAT goes through a sharp threshold (a concept from the random graph category, which is weaker than the phase transition expression) \cite{friedgut1999}. Then the lower and upper bounds of the satisfiable-unsatisfiable transition point were gradually improved in several papers. M{\'{e}}zard {\it et al} applied the cavity method derived from spin-glass research to the CSP and  separated the satisfiable region into two parts, the replica symmetric (RS) part, where the solutions belongs to a single state (cluster), and the one step replica symmetry breaking (1RSB) part, where the solutions belongs to many states \cite{mezard2001}\cite{mezard2002}\cite{mezardsci}. The transition from the RS phase to the 1RSB phase is called the clustering transition.  Further studies using the cavity method found the condensation transition, where the solution space starts to be dominated by a few large clusters \cite{Krzakala}\cite{Montanari}, and the freezing transition, where a linear number of frozen variables (fixed throughout the cluster) arise in almost every cluster \cite{Zdeborova}. It is worth mentioning that inspired by the 1RSB type cavity method, the survey propagation (SP) algorithm \cite{mezard2002}\cite{mezardsci} was proposed and is known to be a highly efficient algorithm when the constraint density is rather close to the satisfiable-unsatisfiable transition.

Solution space structures affect problem hardness and the performance characteristics of different algorithms. Problems approach their maximum hardness when the constraint density approaches the satisfiable-unsatisfiable transition. Experiments suggest that local Monte Carlo Markov chain strategies are effective up to the clustering transition and that belief propagation (BP) is effective up to the condensation transition \cite{Krzakala}. It was stated in \cite{Braunstein} that the intuitive assumption supporting the SP validity is that many clusters exist in the solution space, whereas for BP, it is assumed that most solutions belong to one cluster. Additionally, many researchers believe that the freezing transition is a pivotal transition for algorithm validity.

The random CSP model, also known as the random CSP instances generator, is proposed to enrich the study of the CSP. The initial proposed random models \cite{Gent}\cite{Smith} are called models A, B, C, and D, where the constraint scope size and domain size are fixed. Unfortunately for these models, it was proved by Achlioptas {\it et al} \cite{ach97} that the generated instances suffer trivial unsatisfiability, that is, almost all instances are unsatisfiable when the number of variables is large. To overcome this flaw, many alternative models have been proposed. One technique is to incorporate a special combinatorial structure on constraints and ensure that the generated instance has certain consistency properties \cite{Gent}\cite{Gao07}. Another technique is to  change the scales of the parameters, including the size of the domain and the length of the constraint scope \cite{smith2001,frize,xu2000,fan2011,fan2012}. In this work, we study the d-k-CSP of Ref. \cite{fan2012}, which shows a varying constraint scope length and a varying domain size.

Model RB, a special case of model d-k-CSP, also plays a significant role in computer science. In Ref. \cite{zhao2012}, Bethe free entropy was studied, and it was suggested that the RS solution should always be stable locally; thus, the condensation transition should be absent in this model. In article \cite{xuwei}, we indeed prove that the clustering phase exists and persists until the satisfiable-unsatisfiable transition point, so the condensation phase do not exist. The solution space of model RB is quite different from that of other CSPs such as K-SAT or Coloring, where the condensation transition is observed \cite{Krzakala}. The method applied in \cite{xuwei} was initiated by
M{\'e}zard {\it et al}, who studied the clustering phenomenon by counting the numbers of solution pairs at different distances \cite{MMZ2005}; and developed by Achlioptas {\it et al}, who proved that the clustering phase exists on K-SAT \cite{Achlioptas1,Achlioptas2,Achlioptas3,Achlioptas4}.

In this article, we study the solution space structure of model d-k-CSP. Using the same method as in \cite{xuwei}, we proof that the clustering phase exists before the satisfiable-unsatisfiable transition. There are many well-separated cluster-regions (a cluster-region is a union of some clusters), and the diameter of a cluster decreases with r. Marginals obtained from
Bethe-Peierls approximation show that the clusters distribute widely in the solution space. It concludes that when $r$ is below and close to 1, the solution space contains many widely distributed well-separated small clusters. When r is below and
close to 1, the problem is hard to solve, so the above structure should be the reason of the hardness. As d-k-CSP is an important model of the CSP, our result will provide a further understanding of the relation between solution space structure and complexity.

The rest of the paper is organized as follows: we define the model in Sec. \ref{sec:model}, describe the method to show clusters in Sec. \ref{sec:method}, then prove the clustering phenomenon in Sec. \ref{sec:pair}, \ref{sec:num}, \ref{sec:cluster}, then a special case is discussed in Sec. \ref{sec:spe}, and marginals of variables and distribution of clusters are studied by physical method in Sec.\ref{sec:bp}.

\section{Model d-k-CSP and definitions}\label{sec:model}
Fan, Shen and Xu (2012) proposed a general model of a random CSP with varying constraint scope length and varying domain size, called model d-k-CSP \cite{fan2012}. An instance of model d-k-CSP is composed of a set of variables $V = {x_1, x_2, . . . , x_n}$ and a set of constraints
$C={C_1,C_2, . . . ,C_t}$, where $n,t$ is the number of variables and constraints. Each variable $x_i (i = 1,...,n)$ can only be assigned a value from its domain $D=\{1,2, ...,d\}$, where $d$ is function of $n$, and $d(n)\geq 2$. Each constraint $C_i (i = 1,...,t)$ is a pair $(X_i , R_i )$, where $X_i = (x_{i1}, x_{i2}, . . . , x_{ik} )\subseteq V$ is the constraint scope, $k = k(n) \geq 2$ is the constraint scope length, and $R_i \subseteq D_{i1}\times D_{i2}\times ...\times D_{ik}$ is a set of compatible tuples of values. A d-k-CSP instance I is generated by the following two steps:
\begin{enumerate}
\item Select $t$ constraints randomly with repetition. Each constraint is formed by
selecting $k$ of the $n$ variables randomly without repetition.

\item For each constraint, select $q = (1-p)d^k$ compatible tuples of values randomly without repetition, where $0 < p < 1$ is a
constant.
\end{enumerate}

A constraint $C_i=(X_i , R_i )$ is said to be satisfied by an assignment $\sigma \in D^n$ if the assigned values of $X_i$ are in $R_i$. An assignment $\sigma$ is a solution if it satisfies all constraints. A CSP instance is called satisfiable if there are solutions, and called unsatisfiable otherwise. Fan {\it et al} \cite{fan2012} proved that model d-k-CSP has an satisfiable-unsatisfiable transition: assume that $r > 0$ and $0 < p < 1$ are constants, $t = r\frac{ n \ln d}{-\ln (1-p)}$, $\lim_{b\rightarrow \infty}{\frac{1}{d}}$ exists, $k\geq \frac{1}{1-p}$ and there is a positive real number $\epsilon$ such that $k\ln d \geq (1+\epsilon)\ln n$; then,
$$
\lim_{n\rightarrow \infty }{Pr [\text{I is satisfiable}] =
\begin{cases}
1 &r < 1,\\
0 &r > 1.
\end{cases}}
$$
As part of the proof, they found that in the satisfiable phase $r<1$,
\begin{eqnarray}\label{one}
\lim_{n\to\infty}\frac{(\mathbb{E}(X))^2}{\mathbb{E}(X^2)}= 1.
\end{eqnarray}

\begin{figure}[H]
  \centering
  \includegraphics[width=0.8\columnwidth]{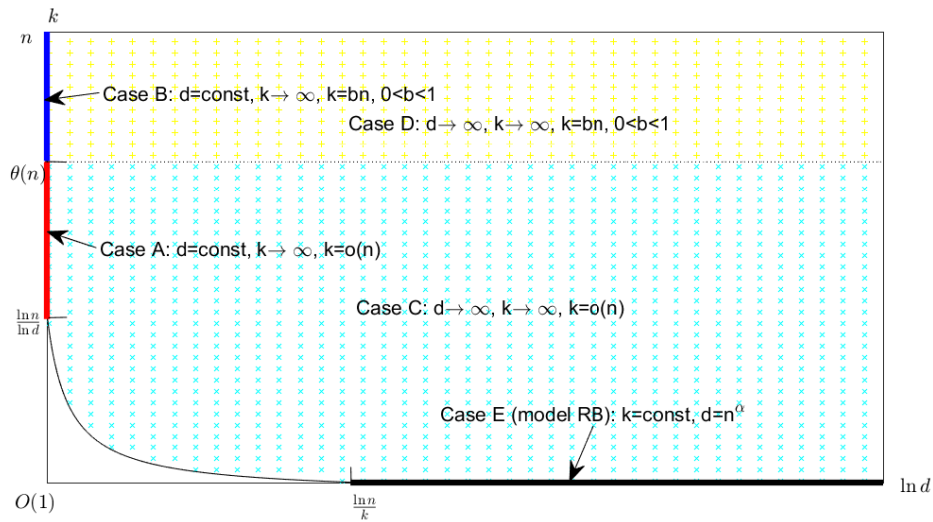}
\caption{The cases in the coordinate system of $\ln d$ and $k$. On the $k$ axis, there are Case A and Case B. On the $\ln d$ axis, there is model RB. In the shadowed area there are Case C and Case D.}\label{sorttu}
\end{figure}

 We sort the model into different cases based on the speeds in which the domain size $k(n)$ and/or the length of constraint scope $d(n)$ grow with the number of variables $n$. First of all the conditions in Theorem 2.1 of \cite{fan2012} should be satisfied, in order to guarantee the existence of the satisfiable-unsatisfiable transition. Then we state other conditions of Case A, B, C, D and model RB in Table \ref{sorttable}. Taking case A for an example, in case A the parameter $d(n)$ is a constant, $k(n)$ tends to infinity, and $k(n)$ is an infinitesimal of higher order than $n$. In Fig. \ref{sorttu}, we show the conditions of the cases in the coordinate system of $\ln d$ and $k$.
\begin{table}[H]
\centering
\begin{tabular}{|c|c|c|c|c|}
\hline
The cases & Other conditions\\
\hline
Case A & $d(n)=const, k(n)\rightarrow \infty, k(n)=o(n)$  \\
\hline
Case B & $d(n)=const, k(n)\rightarrow \infty, k(n)=bn, 0<b<1$\\
\hline
Case C & $d(n)\rightarrow \infty, k(n)\rightarrow \infty, k(n)=o(n)$ \\
\hline
Case D & $d(n)\rightarrow \infty, k(n)\rightarrow \infty, k(n)=bn, 0<b<1$\\
\hline
Model RB \cite{xuwei}&  $k(n)=const, d(n)=n^\alpha$\\
\hline
\end{tabular}\caption{Other conditions of different cases.}\label{sorttable}
\end{table}

A solution pair is a sequence of two assignment solutions. The (Hamming) distance between two solutions is the number of variables which have different values in the two solutions. For a d-k-CSP instance, the number of solution pairs at distance $x (nx=1,2,...,n)$ is denoted by $Z(x)$, and $\mathbb{E}(Z(x))$ is its expectation in the model. {\it Cluster} is the connected component in the solution space,
where every pair of solutions are considered
to be adjacent if they are at distance 1. {\it Cluster-region} is a union of some clusters. The {\it distance between two cluster-regions} is the minimum distance of their solution pairs not belonging to the same cluster-region. The {\it diameter} of a cluster-region is the maximum distance of two solutions in the cluster-region. The {\it clustering phase} describes the phase where the solution space breaks apart into an exponential number of well-separated clusters, with each cluster containing a sub-exponential number of solutions. The {\it condensation phase} describes the phase where a finite number of clusters contain almost all of the solutions, which is different from the clustering phase.

\section{Method to show clusters}\label{sec:method}

The method used in this work is based on the distances among the solutions \cite{xuwei,MMZ2005,Achlioptas1,Achlioptas2,Achlioptas3,Achlioptas4}. If solution-pairs at distances between $\alpha n$ and $\beta n$ do not exist, the solution space can be split into cluster-regions, with the cluster-region diameter smaller than $\alpha n$ and the distance among cluster-regions larger than $\beta n-\alpha n$. With this method, to obtain the clustering properties there are two steps left.

Firstly, we should determine $\alpha$ and $\beta$ such that w.h.p. solution-pairs at distances between $\alpha n$ and $\beta n$ do not exist, where ``w.h.p.'' means the probability of an event tends to 1 as $n\rightarrow \infty$. For this purpose, we need only to prove that there is a positive $\delta$ such that $\sup_{n\rightarrow \infty}f_0(x)<-\delta$ for $\alpha<x<\beta$, where the $f_0(x)$ is defined by $\ln \mathbb{E}(Z(x))/n$. If this is satisfied, by the first moment method, we have
$$\sup_{n\rightarrow \infty}P\left(\sum_{xn=\alpha n}^{\beta n}{Z(x)}>0\right)\leq\sup_{n\rightarrow \infty}\mathbb{E}\left(\sum_{xn=\alpha n}^{\beta n}{Z(x)}\right)=\sup_{n\rightarrow \infty}\sum_{xn=\alpha n}^{\beta n}{e^{nf_0(x)}}\leq\sup_{n\rightarrow \infty}ne^{-\delta n}= 0,$$
which means w.h.p. solution-pairs at distances between $\alpha n$ and $\beta n$ do not exist.
Or we need only to prove that $\sup_{n\rightarrow \infty}g_0(x)<-\delta$ for $\alpha<x<\beta$, where $g_0(x)$ is defined by $\ln \mathbb{E}(Z(x))/(n\ln d)$.

Secondly, we should estimate the number of cluster-regions and the number of solutions in each cluster-region. Let $l=max_{0< xn\leq \alpha n}\mathbb{E}(Z(x))$, by the first moment method and the definition of $l$, we obtain
\begin{eqnarray*}
P(\sum_{xn=1}^{an}{Z(x)}\geq n^2l)\leq \frac{\sum_{xn=1}^{an}{\mathbb{E}(Z(x))}}{n^2l}\leq \frac{nl}{n^2l}\leq \frac{1}{n}\rightarrow 0,
\end{eqnarray*}
 that is, w.h.p., the number of solution-pairs in each cluster-region is smaller than $n^2l$. Then, the number of solutions in each cluster-region is smaller than $nl^{0.5}$.
We can give a lower bound to the number of all solutions; here, using the Paley-Zigmund inequality and Eqn. \ref{one}, we have
\begin{eqnarray*}
P[ X>\frac{1}{n}\mathbb{E}(X) ] \geq& \frac{\left(\mathbb{E}(X)-\frac{1}{n}\mathbb{E}(X)\right)^2}{\mathbb{E}(X^2)}\rightarrow&(1-\frac{1}{n})^2\rightarrow 1.
\end{eqnarray*}
The number of cluster-regions must be larger than the lower bound of the number of all solutions divided by the upper bound of the number of solutions in each cluster-region. It is to say that the number of cluster-regions is larger than
\begin{eqnarray}\label{eq:second_moment}
\frac{\frac{1}{n}\mathbb{E}(X)}{nl^{0.5}}=\frac{\frac{1}{n}d^n(1-p)^t}{nl^{0.5}}.
\end{eqnarray}
To give a lower bound to Eqn. \ref{eq:second_moment} in the limit condition where $n\rightarrow \infty$, we need only give an upper bound to $\sup_{n\rightarrow \infty}l$.

To summarize:
\begin{itemize}
\item Given $\delta>0$, we should find $\alpha$ and $\beta$ that for $\alpha<x<\beta$, $\sup_{n\rightarrow \infty}f_0(x)<-\delta$ or $\sup_{n\rightarrow \infty}g_0(x)<-\delta$.

\item We should give an upper bound to $\sup_{n\rightarrow \infty}l$ to estimate the number of cluster-regions.
\end{itemize}

\section{study the number of solution-pairs and find $\alpha, \beta$}\label{sec:pair}

The expression for $\mathbb{E}(X)$ should be given first. Using the same analysis as in Eqs. (8) and (9) in article \cite{xu2000}, we have
\begin{eqnarray}\label{eq:ezx}
\mathbb{E}(Z(x))=d^nC^{nx}_{n}(d-1)^{nx}\left[\frac{C_{d^k-1}^{q}}{C_{d^k}^{q}}\sigma(x)+\frac{C_{d^k-2}^{q}}{C_{d^k}^{q}}\left(1-\sigma(x)\right)\right]^t,
\end{eqnarray}
where $\sigma(x)=C^k_{n-nx}/C^k_{n}$, and $C_{i}^{j}$ takes the value of 0 by definition if $i<j$. Parameters $n, d, k, q,$ and $t$ are all from the definition of model d-k-CSP, and $d, k,$ and $x$ are actually functions of $n$, that is, $d=d(n),k=k(n),x=x(n)$. We study the limit when $n\rightarrow \infty$.
By simplification and asymptotic estimation, and because $d^k\rightarrow 0$ in all the cases that we study here, we obtain
\begin{center}
$\frac{C_{d^k-1}^{q}}{C_{d^k}^{q}}=1-p$, $\frac{C_{d^k-2}^{q}}{C_{d^k}^{q}}=(1-p)^2+O(d^{-k})\rightarrow (1-p)^2$.
\end{center}
In the following of this section, we focus on the signs of $f_0(x)=\ln{(\mathbb{E}(Z(x)))}/n$ and $g_0(x)=\ln{(\mathbb{E}(Z(x)))}/(n\ln d)$.
When $d=const$, $f_0(x)$ tends to a finite value, and
\begin{align*}
f_0(x)\rightarrow A(x)+B(x),
 \end{align*}
where $$A(x)=1/n(\ln C_{n}^{nx}+nx\ln(d-1)),$$ $$B(x)=\ln d-r\ln d+r\frac{\ln d}{-\ln(1-p)}\ln\left(1-p+p\sigma(x)\right).$$
There is a $\ln C_n^{nx}$ in the above formula, and note that when $0<x<1$ is a constant, by the Stirling formula, as $n\rightarrow \infty$,
\begin{eqnarray*}
\ln C_n^{nx}\rightarrow -n\ln(x^x(1-x)^{1-x}).
\end{eqnarray*}
When $d\rightarrow \infty$, we study $g_0(x)$ instead because it tends to a finite value, and
\begin{align*}
g_0(x)\rightarrow C(x)+D(x),
\end{align*}
where $$C(x)=x,$$ $$D(x)=1-r+r\frac{1}{-\ln(1-p)}\ln\left(1-p+p\sigma(x)\right).$$

Note that parameter $d\geq 2$ and $0\leq\sigma(x)\leq1$, so we have $A(x)$ are positive and $\ln(1-p)\leq\ln\left(1-p+p\sigma(x)\right)\leq0$. Therefore, when $r<0.5$, we have $$A(x)+B(x)>\ln d-2r\ln d>0,$$ $$\lim_{n\rightarrow \infty}f_0(x)\geq0,$$ and also $$\lim_{n\rightarrow \infty}g_0(x)\geq0.$$
It is to say, when $r<0.5$, we can not find $\alpha$, $\beta$ such that for $\alpha<x<\beta$ $\lim_{n\rightarrow \infty}f_0(x)<0$ or $\lim_{n\rightarrow \infty}g_0(x)<0$.

But when $0.5<r<1$, we can find the $\alpha,\beta$ for Cases A, B, C, and D. For each case we give three statements in the following:

(1) The first statement will be that for $0<x<\alpha$ we have $\inf_{n\rightarrow \infty}f_0(x)\geq0$ or $\inf_{n\rightarrow \infty}g_0(x)\geq0$.

(2) The second statement will be that for $\beta<x<\beta^*$ ($\beta^*$ is some constant value bigger than $\beta$) we have $\inf_{n\rightarrow \infty}f_0(x)\geq0$ or $\inf_{n\rightarrow \infty}g_0(x)\geq0$.

(3) The third statement will be that for $\alpha(1+\epsilon)<x<\beta(1-\epsilon)$ ($\epsilon$ is arbitrarily small) we have $\sup_{n\rightarrow \infty}f_0(x)<-\delta$ or $\sup_{n\rightarrow \infty}g_0(x)<-\delta$ ($\delta$ is positive and depends on $\epsilon$).

\subsection{Case A where $d=const, k\rightarrow \infty$, and $k=o(n)$}

(1) For a constant $a_0>0$ defined in the following, when $x\in(0,a_0/k)$, we have $\inf_{n\rightarrow \infty}f_0(x)\geq0$.

In this case $k\rightarrow \infty$, for any constant $a$, when $xn\in(0,an/k)$, we have $x\rightarrow 0$ and $f_0(x)\rightarrow B(x)$.
If $x=a/k$, where $a=const$, we have
\begin{align*}
\sigma(x)=C^k_{n-nx}/C^k_{n}=\frac{(n-nx)...(n-nx-k+1)}{n...(n-k+1)}<(1-\frac{nx}{n})^k\rightarrow e^{-xk}=e^{-a},
\end{align*}
\begin{align*}
\sigma(x)=C^k_{n-nx}/C^k_{n}>(1-\frac{nx}{n-k+1})^{\frac{n-k+1}{nx}\frac{nx}{n-k+1}k}\rightarrow e^{-\frac{nx}{n-k+1}k}\rightarrow e^{-a},
\end{align*}
and combining the above two inequalities, we obtain $\sigma(x)\rightarrow e^{-a}$. Then, we find $B(a/k)\rightarrow f_1(a)$, where $f_1(a)$ is defined by
\begin{align*}
f_1(a)\triangleq\ln d-r\ln d+r\frac{\ln d}{-\ln(1-p)}\ln\left(1-p+pe^{-a}\right),
\end{align*}
noting that $f_1(a)$ decreases with $a$. Solving equation $f_1(a)=0$ with variable $a$, we obtain the solution
\begin{eqnarray*}
-\ln \frac{(1-p)^{1/r}-(1-p)^2}{1-p-(1-p)^2}\triangleq a_0,
\end{eqnarray*}
which is a positive constant when $0.5<r<1$. Function $f_0(x)\rightarrow B(x)$, $B(x)$ is a decreasing function, and $B(a_0/k)\rightarrow f_1(a_0)=0$; therefore, we find that when $x<a_0/k$, $\inf_{n\rightarrow \infty}f_0(x)\geq0$.

(2) For constants $b_0$ and $b_1$ defined in the following, when $x\in (b_0, b_1)$, we have $\inf_{n\rightarrow \infty}f_0(x)\geq0$.

\begin{figure}[H]
  \centering
  \includegraphics[width=0.6\columnwidth]{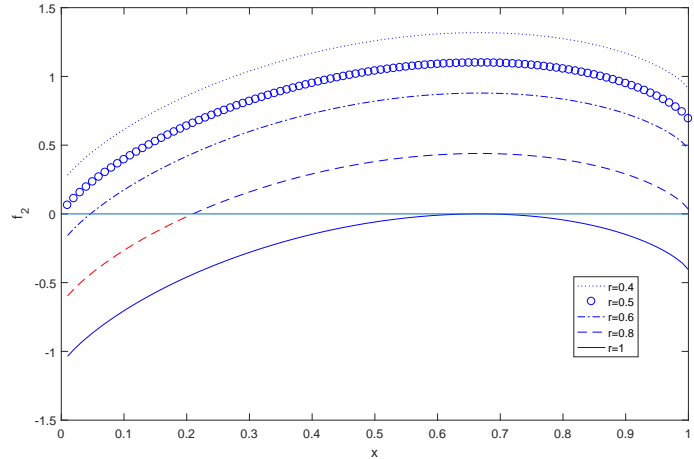}
  \caption{Function $f_2$ for $d=3$ and $r=0.4,0.5,0.6,0.8,1$ from top to bottom. There is a region satisfying $f_2(x)<0$ when $r>0.5$. Taking $r=0.8$ as an example, in the region $x\in (0, 0.2)$, $f_2(x)<0$ (plotted in red). }\label{f2tu}
\end{figure}

If $x$ is a positive constant, then $\sigma (x)=C^k_{n-nx}/C^k_{n}\rightarrow 0$, and
\begin{eqnarray}\label{f2x}
f_0(x)\rightarrow\ln d-\ln (x^x(1-x)^{1-x})+x\ln(d-1)-2r\ln d\triangleq f_2(x).
\end{eqnarray}
We can draw a picture of function $f_2(x)$, as shown in Fig \ref{f2tu}.
The first- and second-order derivatives of $f_2(x)$ are
\begin{eqnarray*}
f'_2(x)=-\ln x+\ln (1-x)+\ln (d-1),
\end{eqnarray*}
\begin{eqnarray*}
f''_2(x)=-\frac{1}{x}-\frac{1}{1-x}<0.
\end{eqnarray*}
$f_2(x)$ is a concave function. Let $f'_2(x)=0$; we then obtain $x=\frac{d-1}{d}$. Then, $f_2(x)$ achieves its maximum value at $$x=\frac{d-1}{d}\triangleq b_1,$$ and the maximum value is $$f_2(x=\frac{d-1}{d})=2(1-r)\ln d.$$ When $0.5<r<1$,  $$f_2(x=\frac{d-1}{d})=2(1-r)\ln d>0,$$ $$f_2(x=0)=(1-2r)\ln d<0,$$ so equation $f_2(x)=0$ has one solution in region $(0,\frac{d-1}{d})$, denoted by $b_0$.
Additionally, there is a region $(b_0, b_1)$ where $f_2(x)>0$, and as $f_0(x)\rightarrow f_2(x)$, we obtain $\inf_{n\rightarrow \infty}f_0(x)>0$.

(3) For arbitrarily small positive number $\epsilon$, when $x\in ((a_0+\epsilon)/k,b_0-\epsilon)$, there exists $\delta>0$ s.t. $\sup_{n\rightarrow \infty}f_0(x)<-\delta$.

$A(x)$ is very small when $x$ is a very small constant, so for $-\frac{1}{2}f_1(a_0+\epsilon)$, which is a positive number, there exists a constant $x_0$ such that $x_0<b_0-\epsilon$, $x_0<1/2$  and  $$A(x_0)<-\frac{1}{2}f_1(a_0+\epsilon).$$
Based on this $x_0$, we let $$((a_0+\epsilon)/k,b_0-\epsilon)=((a_0+\epsilon)/k,x_0]\cup (x_0,b_0-\epsilon).$$
In the first range $((a_0+\epsilon)/k,x_0]$, $A(x)$ increases, and therefore,  $A(x)<-\frac{1}{2}f_1(a_0+\epsilon)$;
$B(x)$ decreases, so $B(x)<B((a_0+\epsilon)/k)\rightarrow f_1(a_0+\epsilon)$.
We have in the first range $$\sup_{n\rightarrow \infty}f_0(x)\leq-\frac{1}{2}f_1(a_0+\epsilon)+f_1(a_0+\epsilon)=\frac{1}{2}f_1(a_0+\epsilon).$$
In the second range $(x_0,b_0-\epsilon)$, we have $f_0(x)\rightarrow f_2(x)<f_2(b_0-\epsilon)$, and therefore $$\sup_{n\rightarrow \infty}f_0(x)\leq f_2(b_0-\epsilon).$$
Let $-\delta=max(\frac{1}{2}f_1(a_0+\epsilon), f_2(b_0-\epsilon))$; we have for $x\in ((a_0+\epsilon)/k,b_0-\epsilon)$, $\sup_{n\rightarrow \infty}f_0<-\delta$.

\subsection{Case B where $d=const, k\rightarrow \infty, k=bn$, and $0<b<1$}

(1) For a constant $a_1>0$ defined in the following, when $x\in(0,a_1/n)$, we have $\inf_{n\rightarrow \infty}f_0(x)\geq0$.

If $x=a/n$, where $a$ is a positive constant integer, we have
\begin{eqnarray*}
\sigma(x=a/n)=\frac{C^{bn}_{n-a}}{C^{bn}_n}=\frac{(n-a)...(n-a-bn+1)}{n...(n-bn+1)}=\frac{(n-bn)...(n-bn-a+1)}{n...(n-a+1)}
\rightarrow (1-b)^a;
\end{eqnarray*}
then, we find $f_0(a/n)\rightarrow f_3(a)$, where $f_3(a)$ is defined by
\begin{align*}
f_3(a)\triangleq \ln d-r\ln d+r\frac{\ln d}{-\ln(1-p)}\ln\left(1-p+p(1-b)^{a}\right).
\end{align*}
Solving $f_3(a)=0$ with variable $a$, we obtain the solution
\begin{eqnarray*}
\frac{\ln((1-p)^\frac{1-r}{r}-1+p)-\ln p}{\ln(1-b)}\triangleq a_1.
\end{eqnarray*}
When $0.5<r<1$, we have $0<\ln((1-p)^\frac{1-r}{r}-1+p)<p$, and then, $a_1$ is a positive constant. When $x<a_1/n$, $f_0(x)\rightarrow B(x)$, $B(x)$ is a decreasing function and $B(a_1/n)\rightarrow f_3(a_1)=0$, so $\inf_{n\rightarrow \infty}f_0(x)\geq0$.

(2) As the same as Case A,  if $x$ is a positive constant, $f_0(x)\rightarrow f_2(x)$. Then, for  $b_0$ and $b_1$, we have $\inf_{n\rightarrow \infty}f_0(x)\geq0$ when $x\in (b_0, b_1)$.

(3) For arbitrarily small positive number $\epsilon$, when $x\in ((a_1+\epsilon)/n,b_0-\epsilon)$, there exists $\delta>0$ s.t. $\sup_{n\rightarrow \infty}f_0(x)<-\delta$.

$-\frac{1}{2}f_3(a_1+\epsilon)$ is a positive number, and same as for case A, there exists a constant $x_1$ such that $x_1<b_0-\epsilon$, $x_1<1/2$  and  $$A(x_1)<-\frac{1}{2}f_3(a_1+\epsilon).$$
Based on this $x_1$, we divide range $((a_1+\epsilon)/n,b_0-\epsilon)$ into $$((a_1+\epsilon)/n,x_1]\cup (x_1,b_0-\epsilon).$$
In $((a_1+\epsilon)/n,x_1]$, $A(x)$ increases, and therefore, $A(x)<-\frac{1}{2}f_3(a_1+\epsilon)$;
$B(x)$ decreases, so $B(x)<B((a_1+\epsilon)/n)\rightarrow f_3(a_1+\epsilon)$.
We have in this range $$\sup_{n\rightarrow \infty}f_0(x)\leq\frac{1}{2}f_3(a_1+\epsilon).$$
In $(x_1,b_0-\epsilon)$, we have $f_0(x)\rightarrow f_2(x)<f_2(b_0-\epsilon)$, and then $\sup_{n\rightarrow \infty}f_0(x)\leq f_2(b_0-\epsilon)$.

Let $-\delta=max(\frac{1}{2}f_3(a_1+\epsilon), f_2(b_0-\epsilon))$; we have for $x\in ((a_1+\epsilon)/n,b_0-\epsilon)$, $\sup_{n\rightarrow \infty}f_0<-\delta$.

\subsection{Case C where $d\rightarrow \infty, k\rightarrow \infty$, and $k=o(n)$}

(1) If $x=a/k$, where $a=const$, we have
$\sigma(x)\rightarrow e^{-a}$,
\begin{align*}
g_0(x)\rightarrow 1-r+r\frac{1}{-\ln(1-p)}\ln\left(1-p+pe^{-a}\right)\triangleq g_1(a).
 \end{align*}

Solving equation $g_1(a)=0$ with variable $a$, the solution  $a_0$ is obtained. Similarly to case A, when $x\in(0,a_0/k)$, we have $\inf_{n\rightarrow \infty}g_0(x)\geq0$.

\begin{figure}[H]
  \centering
  \includegraphics[width=0.6\columnwidth]{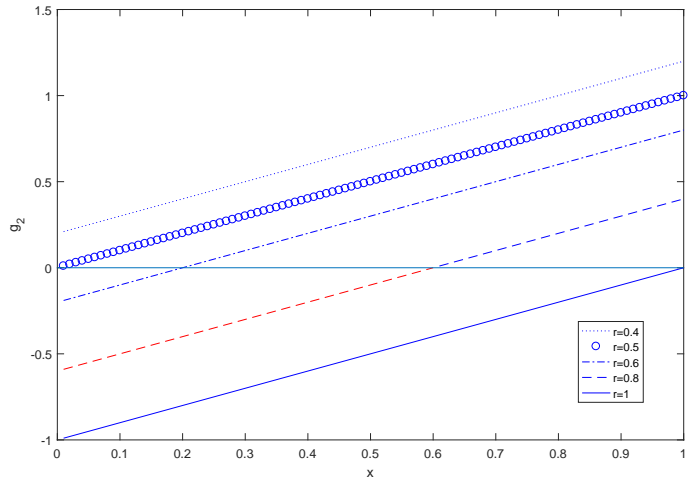}
  \caption{Function $g_2(x)$, where $r=0.4,0.5,0.6,0.8,1$ from top to bottom. There is a region satisfying $g_2(x)<0$ when $r>0.5$. Taking $r=0.8$ for example, in the region $x\in (0, 0.6)$, $g_2(x)<0$ (plotted in red). }\label{g2tu}
\end{figure}

(2) If $x$ is a positive constant, then $\sigma (x)\rightarrow 0$, and
\begin{align}\label{g2x}
g_0(x)\rightarrow 1+x-2r\triangleq g_2(x).
\end{align}
We can draw an illustration of function $g_2$ with variable $x$ in Fig \ref{g2tu}.
Equation $g_2(x)=0$ has one solution, which is $2r-1$. $g_2(x)$ increases, so we let $b_2=2r-1$ and $b_3>b_2$. Then, when $x\in (b_2, b_3)$, $\inf_{n\rightarrow \infty}g_0(x)\geq0$.

(3) Note that C(x) increases and D(x) decreases, then via a similar process, we find that for arbitrarily small positive number $\epsilon$, when $x\in ((a_0+\epsilon)/k,b_2-\epsilon)$, there exists $\delta=max(\frac{1}{2}g_1(a_0+\epsilon), g_2(b_2-\epsilon))$ such that $\sup_{n\rightarrow \infty}g_0(x)<-\delta$.

\subsection{Case D where $d\rightarrow \infty, k\rightarrow \infty, k=bn$, and $0<b<1$}

(1) If $x=a/n$, where $a=const$, we have
$\sigma(x)\rightarrow (1-b)^a$,
\begin{align*}
g_0(x)\rightarrow 1-r+r\frac{1}{-\ln(1-p)}\ln\left(1-p+p(1-b)^a\right)\triangleq g_3(a).
 \end{align*}

Solving equation $g_3(a)=0$ with variable $a$, solution $a_1$ is obtained. Similarly to Case C, when $x\in(0,a_1/n)$, we have $\inf_{n\rightarrow \infty}g_0(x)\geq0$.

(2) As the same as Case C, if $x$ is a positive constant, $g_0(x)\rightarrow g_2(x)$; then, for  $b_2$ and $b_3$, we have  $\inf_{n\rightarrow \infty}g_0(x)\geq0$ when $x\in (b_2, b_3)$.

(3) Similarly to Case C, for arbitrarily small positive number $\epsilon$, when $x\in ((a_1+\epsilon)/n,b_2-\epsilon)$, there exists $\delta=max(\frac{1}{2}g_3(a_1+\epsilon), g_2(b_2-\epsilon))$ such that $\sup_{n\rightarrow \infty}g_0(x)<-\delta$.
~\\

\section{the number of clusters}\label{sec:num}
$l=max_{0< xn\leq \alpha n}\mathbb{E}(Z(x))$, where $\alpha=a_0/k$ in cases A, and C and $\alpha=a_1/n$ in cases B and D. Therefore, for $0< x\leq \alpha$, in cases A, B, C, and D, we have $x\rightarrow 0$, and
\begin{align*}
\ln{(\mathbb{E}(Z(x)))}/(n\ln d)\rightarrow D(x)=1-r+r\frac{1}{-\ln(1-p)}\ln\left(1-p+p\sigma(x)\right))\leq 1-r,
\end{align*}
where the last inequality is because $0 \leq \sigma(x)\leq 1$. By the above inequality, we obtain $$\sup_{n\rightarrow \infty}{l}\leq n^{\ln d-r\ln d}.$$
When $r<1$, we obtain a lower bound for the value of Eqn. \ref{eq:second_moment}, and the number of cluster-regions is larger than $$\frac{\frac{1}{n}d^n(1-p)^t}{nl^{0.5}}\geq \frac{1}{n^2}d^{0.5(1-r)n}.$$ Then the number of cluster-regions increases exponentially with $n$.

\section{Clustering phase of model d-k-CSP}\label{sec:cluster}
The method in Sec. \ref{sec:method} tells us that, if solution-pairs at distance between $\alpha n$ and $\beta n$ do not exist, the clustering phase shows, with the cluster diameter smaller than $\alpha n$ and the distance among clusters larger than $\beta n-\alpha n$. In Sec. \ref{sec:pair} we found such $\alpha, \beta$ for Cases A, B, C and D and $0.5<r<1$. And in Sec. \ref{sec:num} we obtained the number of clusters, which increases exponentially with $n$. We list all those properties in  Table \ref{result}.
\begin{table}[H]
\centering
\begin{tabular}{|c|c|c|c|c|}
\hline
Cases & Clustering range & Cluster diameter $\leq$& Distance among cluster-regions $\geq$& Number of clusters\\
\hline
Case A &  $0.5<r<1$  &$a_0n/k$ & $b_0n-a_0n/k\approx b_0n$ & exponential\\
\hline
Case B & $0.5<r<1$  & $a_1$ & $b_0n-a_1\approx b_0n$ &  exponential\\
\hline
Case C &  $0.5<r<1$ & $a_0n/k$ & $(2r-1)n-a_0n/k\approx (2r-1)n$ &  exponential\\
\hline
Case D & $0.5<r<1$   & $a_1$ & $(2r-1)n-a_1\approx (2r-1)n$ &  exponential\\
\hline
Model RB \cite{xuwei}&  $r_0<r<1$ & $a_2n$ & $(b_2-a_2)n$ &  exponential\\
\hline
\end{tabular}\caption{Details of clustering phenomenon in different cases.}\label{result}
\end{table}

In Table
\ref{result}, $a_0=-\ln \frac{(1-p)^{1/r}-(1-p)^2}{1-p-(1-p)^2}$; $a_1=\frac{\ln((1-p)^\frac{1-r}{r}-1+p)-\ln p}{\ln(1-b)}$; $b_0$ is the solution $x\in(0,\frac{d-1}{d})$ of the function $\ln d-\ln (x^x(1-x)^{1-x})+x\ln(d-1)-2r\ln d=0$. $a_2$ and $b_2$ are the two solutions of equation $\alpha(1+x)+r\ln[(1-p)^2+p(1-p)(1-x)^k]=0$ with variable $x$; and $r_0$ is the smallest value for which this equation has not least a solution in $x\in [0,1]$. The results of model RB are from Ref. \cite{xuwei}. We have the following observations.


\begin{enumerate}[(1)]
\item There is an exponential number of cluster-regions, with each cluster-region containing a
sub-exponential number of solutions.
\item The smallest distances among cluster-regions are $\Theta(n)$ (of the same order of $n$), so the cluster-regions are well-separated.
\item With $r$ approaching 1, the diameter of a cluster-region decreases to a small value.
\item With $r$ approaching 1, the distance among the cluster-regions increases, which also means that the number of cluster-regions decreases because of that some cluster-regions disappear.
\item For fixed $r$, as $k$ increases, the diameter of the cluster decreases. When $k$ is a constant, the biggest diameter is $a_2n$; when $k\rightarrow \infty$ but $k=o(n)$, the biggest diameter is $a_0n/k$; when $k=bn$, with $0<b<1$, it is $a_1$. In summary the biggest diameter is $\Theta(n/k)$.
\item For fixed $r$, comparing cases A, C with cases B, D, we find that $d$ has a strong effect on the distance among the cluster-regions. When $d$ is a constant, the smallest distance is $b_0n$; when $d\rightarrow \infty$, the smallest distance is $(2r-1)n$.
\item Condensation phase does not exist, because when $0.5<r<1$ it is in clustering phase and when $r>1$ it is in unsatisfiable phase.
\end{enumerate}
From the above observations (1-3), when r is below and close to 1, the solution space contains many well-separated small cluster-regions.

\section{A special case where $k=n$}\label{sec:spe}

If $k=n$, for all $xn=1,2,...,n$, we have $\sigma(x)=0$, and then from Eqn. \ref{eq:ezx} we have
\begin{eqnarray*}
\mathbb{E}(Z(x))\rightarrow d^nC^{nx}_{n}(d-1)^{nx}(1-p)^{2t}.
\end{eqnarray*}
Furthermore if $d=const$, we have $f_0(x)\rightarrow f_2(x)$ for $x=1/n,2/n,...,1$, where $f_2(x)$ is defined in Eqn. \ref{f2x} and is shown in Fig \ref{f2tu}. If $d\rightarrow \infty$, we have $g_0(x)\rightarrow g_2(x)$ for $x=1/n,2/n,...,1$, where $g_2(x)$ is defined in Eqn. \ref{g2x} and is shown in Fig \ref{g2tu}. Therefore, when $0.5<r<1$ and $d=const$, w.h.p. solution-pairs at a distance between $1$ and $b_0n$ do not exist;  when $0.5<r<1$ and $d=const$  w.h.p. solution-pairs at a distance between $1$ and $2r-1$ do not exist. Then, this special case has a special solution space structure that the solutions are isolated from each other. If we regard an isolated solution as a cluster, the details of the clustering can be
shown in Table \ref{table3}. Notice that the cluster diameter is 0, which means that each cluster only contains a single solution.

\begin{table}[H]
\centering
\begin{tabular}{|c|c|c|c|c|}
\hline
 & clustering range & cluster diameter & distance among clusters & number of clusters\\
\hline
$k=n$ and $d=const$ & $0.5<r<1$  & $0$ & $b_0n$ &  exponential\\
\hline
$k=n$ and $d\rightarrow \infty$& $0.5<r<1$   & $0$ & $(2r-1)n$ &  exponential\\
\hline
\end{tabular}\caption{Details of clustering phenomenon in the special case where $k=n$.}\label{table3}
\end{table}

\section{marginals of variables and distribution of clusters}\label{sec:bp}
In this section we study marginals of variables and distribution of clusters. For single CSP instance, the marginals of variables can be approximated through a method which is called Bethe-Peierls approximation or Belief Propagation (BP). The Belief Propagation is an iterative ¡®message-passing¡¯ algorithm, where the update rules are that
\begin{align}
\nu_{i\rightarrow a}^{(t+1)}(x_i)\cong\prod_{b\in \partial i\setminus a} \widehat{\nu}_{b\rightarrow i}^{(t)}(x_i),\label{ite1}\\
\widehat{\nu}_{a\rightarrow i}^{(t)}(x_i)\cong\sum_{\underline{x}_{\partial a\setminus i}}\psi_a(\underline{x}_{\partial a})\prod_{k\in \partial a\setminus i} \nu_{k\rightarrow a}^{(t)}(x_k),\label{ite2}
\end{align}
where $\nu_{i\rightarrow a}^{(t)}$, $\widehat{\nu}_{a\rightarrow i}^{(t)}$ are the messages between variable node $i$ and its adjacent constraint node $a$ at step $t$. The symbol $\cong$ denotes ¡®equality up to a normalization¡¯, because BP messages are understood to be probability distributions. $\psi_a$ is equal to 1 if constraint $a$ is satisfied by $\underline{x}_{\partial a}$, and is equal to 0 otherwise. After the iteration converges, let $\nu_{i\rightarrow a}^{(t)}$ converging to $\nu_{i\rightarrow a}$ and $\widehat{\nu}_{a\rightarrow i}^{(t)}$ converging to $\widehat{\nu}_{a\rightarrow i}$, then the marginal probability that variable $i$ equals to $x_i$ is
\begin{align}
\nu_{i}(x_i)\cong\prod_{b\in \partial i} \widehat{\nu}_{b\rightarrow i}(x_i).\label{margin}
\end{align}

\begin{figure}[H]
  \centering
  \includegraphics[width=0.6\columnwidth]{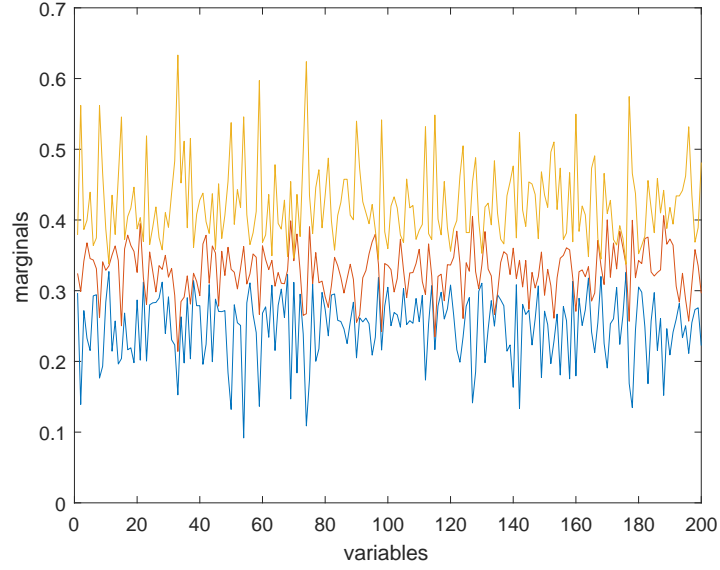}
  \caption{Marginals of 200 variables where we set $n=200, d=3, k=5, p=0.5, r=0.9$. The top line represents the biggest marginal probability for each variable; the second line from above represents the second biggest marginal probaility for each variable; and so on. The averages of the values in different lines are 0.4243,    0.3274,    0.2483.}\label{mar1}
\end{figure}

\begin{figure}[H]
  \centering
  \includegraphics[width=0.6\columnwidth]{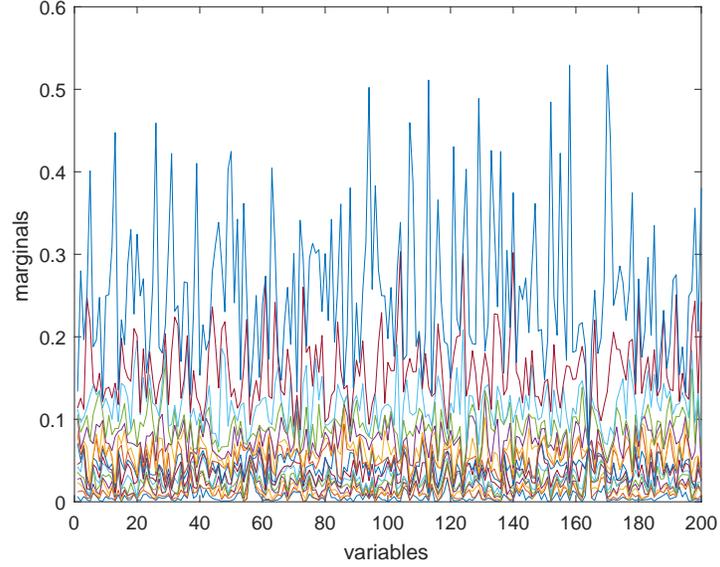}
  \caption{Marginals of 200 variables where we set $n=200, d=15, k=2, p=0.5, r=0.9$. The top line represents the biggest marginal probability for each variable; the second line from above represents the second biggest marginal probability for each variable; and so on. The averages of the values in different lines are 0.2568, 0.1595, 0.1178, 0.0917, 0.0759, 0.0622, 0.0525, 0.0438, 0.0363, 0.0295, 0.0237, 0.0189, 0.0151, 0.0104, 0.0059.}\label{mar2}
\end{figure}

Here we take two instances for examples and estimate the marginals by Belief Propagation. In the first instance we set $n=200, d=3, k=5, p=0.5, r=0.9$, and the result in Fig. \ref{mar1} shows that the marginals are fairly uniform. In the other instance we set $n=200, d=15, k=2, p=0.5, r=0.9$, and the result in Fig. \ref{mar2} shows that almost all marginals are positive and many of them are around the average value 0.067. To sum up, the examinations show that the marginals are uniform to a certain extent.

\begin{figure}[H]
  \centering
  \includegraphics[width=0.6\columnwidth]{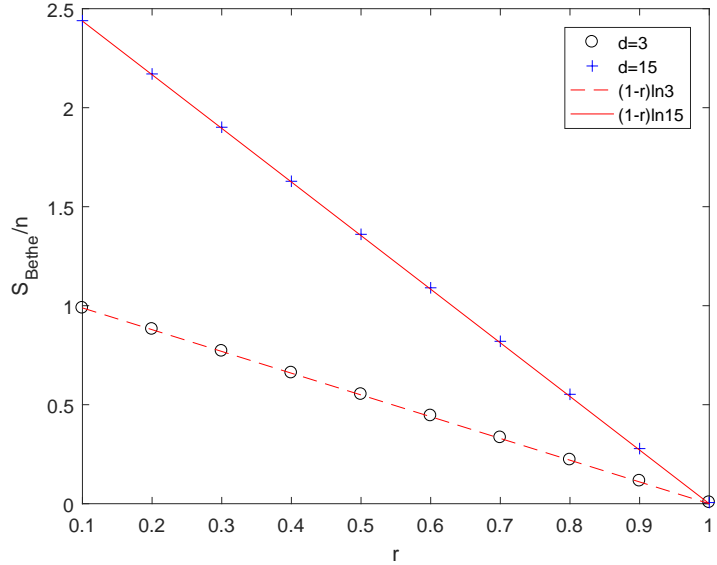}
  \caption{Bethe free entropy $S_{Bethe}/n$, where $r=0.1,0.2,...,1$, $n=200$, $p=0.5$. It shows that $s$ coincides with $\ln [\mathbb{E}(X)]/n=(1-r)\ln d$. Each point is averaged over 30 instances.}\label{shang}
\end{figure}

We now study on whether BP gives the correct marginals. Bethe free entropy is a function of converged messages $\nu_{i\rightarrow a}$s and $\widehat{\nu}_{a\rightarrow i}$s,
\begin{align*}
S_{Bethe}=\sum_aS_a+\sum_iS_i-\sum_{(i,a)}S_{ia},
\end{align*}
where
\begin{align*}
S_a=\log\left[\sum_{\underline{x}_{\partial a}}\psi_a(\underline{x}_{\partial a})\prod_{i\in \partial a} \nu_{i\rightarrow a}(x_i)\right],\\
S_i=\log\left[\sum_{x_i}\prod_{b\in \partial i} \widehat{\nu}_{b\rightarrow i}(x_i)\right],\\
S_{ia}=\log\left[\sum_{x_i}\nu_{i\rightarrow a}(x_i)\widehat{\nu}_{a\rightarrow i}(x_i)\right].
\end{align*}
Here we take $n=200, d=3, k=4, p=0.5$ and $n=200, d=15, k=2, p=0.5$ for examples, and calculate the Bethe free entropy. In Fig. \ref{shang}, it shows that $S_{Bethe}/n$ coincides with the logarithm of the average number of solutions divided by $n$ (annealed entropy density), which is $\ln [\mathbb{E}(X)]/n=(1-r)\ln d$. This indicates that the replica symmetry solution should always be stable locally, and that BP gives the correct marginals.

\begin{figure}[H]
  \centering
  \includegraphics[width=0.6\columnwidth]{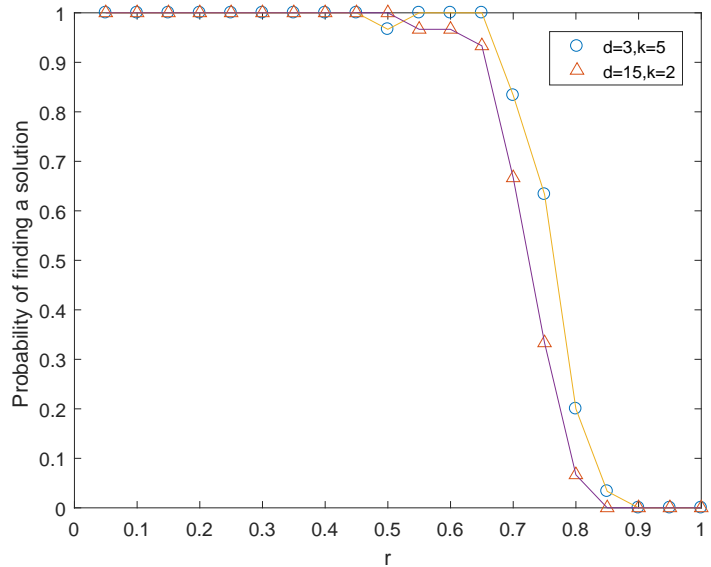}
  \caption{Probability of finding a solution by BP decimation algorithm, where $r=0.1,0.2,...,1$, $n=200$, $p=0.5$. Each point is averaged over 30 instances.}\label{bp}
\end{figure}

We use the BP decimation algorithm on the same instances as in Fig. \ref{shang}. At each step, BP decimation algorithm performs BP iteration (\ref{ite1}) and (\ref{ite2}), estimates marginals by (\ref{margin}), finds the most polarized variable (the one which has the largest marginal probability), fixes the variable to its most possible value, and reduces the problem. If BP iteration always converges and the assignment of the $n$ variables is a solution, the solving process is successful, otherwise failed. Fig. \ref{bp} shows that the algorithm works at about $r<0.7$. This suggests that BP sufficiently reveals the distribution of solutions and gives correct marginals at least in the region $r<0.7$.

Both the research on Bethe free entropy and BP decimation algorithm suggest that BP gives the correct marginals. From the examination of estimation of marginals (such as in Fig. \ref{mar1} and Fig. \ref{mar2}), we find that the marginals are uniform to a certain extent, which means that the clusters distribute widely in the solution space.
\section{Conclusion}
Model d-k-CSP is a standard prototype of Constraint Satisfaction Problem (CSP), where the domain size $d$ and/or the length of constraint scope $k$ grow with the number of variables $n$. Firstly we use a mathematical method to show that, before the satisfiable-unsatisfiable transition, the solution space shatters into an exponential number of well-separated cluster-regions, and the diameter of a cluster decreases with $r$. Secondly physical method shows that the clusters distribute widely in the solution space. So when $r$ is below and close to 1, the solution space contains many widely distributed well-separated small clusters. When $r$ is below and close to 1, the instances are hard to solve, so this solution space structure will lead to high problem hardness.
\section{Acknowledgments}
Project supported by the Fundamental Research Funds for the Central Universities (No. FRF-TP-16-065A1), and the National Natural Science Foundation of China (No. 11801028 and No. 61702019).


\begin{thebibliography}{22}
\expandafter\ifx\csname natexlab\endcsname\relax\def\natexlab#1{#1}\fi
\expandafter\ifx\csname bibnamefont\endcsname\relax
  \def\bibnamefont#1{#1}\fi
\expandafter\ifx\csname bibfnamefont\endcsname\relax
  \def\bibfnamefont#1{#1}\fi
\expandafter\ifx\csname citenamefont\endcsname\relax
  \def\citenamefont#1{#1}\fi
\expandafter\ifx\csname url\endcsname\relax
  \def\url#1{\texttt{#1}}\fi
\expandafter\ifx\csname urlprefix\endcsname\relax\def\urlprefix{URL }\fi
\providecommand{\bibinfo}[2]{#2}
\providecommand{\eprint}[2][]{\url{#2}}




\bibitem{chee}
  \bibinfo{author}{\bibfnamefont{Cheeseman}~\bibnamefont{P}},
  \bibinfo{author}{\bibfnamefont{Kanefsky}~\bibnamefont{B}},
  \bibnamefont{and}
  \bibinfo{author}{\bibfnamefont{Taylor}~\bibnamefont{W~M}}
  \bibinfo{year}{1991}
  in\emph{\bibinfo{booktitle}{Proc. IJCAI}}
 pp
  \bibinfo{pages}{331-337}

\bibitem{friedgut1999}
  \bibinfo{author}{\bibfnamefont{Friedgut~E}},
  \bibnamefont{and}
  \bibinfo{author}{\bibfnamefont{Bourgain}~\bibnamefont{J}}
  \bibinfo{year}{1999}
 \emph{ \bibinfo{journal}{J. Am. Math. Soc.} }\textbf{\bibinfo{volume}{12}}
  \bibinfo{pages}{1017-1054}

\bibitem{mezard2001}
\bibinfo{author}{\bibfnamefont{M{\'{e}}zard}~\bibnamefont{M}}
  \bibnamefont{and} \bibinfo{author}{\bibfnamefont{Parisi}~\bibnamefont{G}}
  \bibinfo{year}{2001}
 \emph{  \bibinfo{journal}{Eur. Phys. J. B} }\textbf{\bibinfo{volume}{20}}
  \bibinfo{pages}{217-233}

\bibitem{mezard2002}
\bibinfo{author}{\bibfnamefont{M{\'{e}}zard}~\bibnamefont{M}}
  \bibnamefont{and} \bibinfo{author}{\bibfnamefont{Zecchina}~\bibnamefont{R}}
  \bibinfo{year}{2002}
 \emph{  \bibinfo{journal}{Phys. Rev. E}} \textbf{\bibinfo{volume}{66}}
  \bibinfo{pages}{056126}

\bibitem{mezardsci}
\bibinfo{author}{\bibfnamefont{M{\'{e}}zard}~\bibnamefont{M}},
  \bibinfo{author}{\bibfnamefont{Parisi}~\bibnamefont{G}},  \bibnamefont{and}
  \bibinfo{author}{\bibfnamefont{Zecchina}~\bibnamefont{R}}
  \bibinfo{year}{2002}
 \emph{  \bibinfo{journal}{Science}} \textbf{\bibinfo{volume}{297}}
  \bibinfo{pages}{812}

\bibitem{Krzakala}
\bibinfo{author}{\bibfnamefont{Krzakala}~\bibnamefont{F}},
  \bibinfo{author}{\bibfnamefont{Montanari}~\bibnamefont{A}},
  \bibinfo{author}{\bibfnamefont{Ricci-Tersenghi}~\bibnamefont{{F}}},
  \bibinfo{author}{\bibfnamefont{Semerjian}~\bibnamefont{G}},
  \bibnamefont{and}
  \bibinfo{author}{\bibfnamefont{Zdeborova}~\bibnamefont{L}} \bibinfo{year}{2007} in
  \emph{\bibinfo{booktitle}{Proc. Natl. Acad. Sci. USA}}
  vol.~\bibinfo{volume}{104}  pp
  \bibinfo{pages}{10318-10323}

\bibitem{Montanari}
\bibinfo{author}{\bibfnamefont{Montanari} \bibnamefont{A}},
  \bibinfo{author}{\bibfnamefont{Ricci-Tersenghi}~\bibnamefont{F}}, \bibnamefont{and}
  \bibinfo{author}{\bibfnamefont{Semerjian}~\bibnamefont{G}}
  \bibinfo{year}{2008}
 \emph{  \bibinfo{journal}{J. Stat. Mech.: Theory Exp.}}
    \bibinfo{pages}{P04004}

\bibitem{Zdeborova}
  \bibinfo{author}{\bibfnamefont{Zdeborova}~\bibnamefont{L}},
  \bibnamefont{and}
  \bibinfo{author}{\bibfnamefont{Krzakala}~\bibnamefont{F}}
  \bibinfo{year}{2007}
  \emph{\bibinfo{booktitle}{Phys. Rev. E}}
  ~\bibinfo{volume}{76}
  \bibinfo{pages}{031131}

\bibitem{Braunstein}
\bibinfo{author}{\bibfnamefont{Braunstein}~\bibnamefont{A}},
 \bibinfo{author}{\bibfnamefont{M{\'{e}}zard}~\bibnamefont{M}},
 \bibnamefont{and}
  \bibinfo{author}{\bibfnamefont{Zecchina}~\bibnamefont{R}}
  \bibinfo{year}{2002}
  \emph{ \bibinfo{journal}{Rand. Struct. Algorithms} }\textbf{\bibinfo{volume}{27}}
  \bibinfo{pages}{201-226}

\bibitem{Gent}
\bibinfo{author}{\bibfnamefont{Gent}~\bibnamefont{{I~P}}},
\bibinfo{author}{\bibfnamefont{Macintyre}~\bibnamefont{{E}}},
\bibinfo{author}{\bibfnamefont{Prosser}~\bibnamefont{{P}}},
\bibinfo{author}{\bibfnamefont{Smith}~\bibnamefont{{B~M}}},
 \bibnamefont{and}
  \bibinfo{author}{\bibfnamefont{Walsh}~\bibnamefont{T}}
  \bibinfo{year}{2001}
\emph{  \bibinfo{journal}{Constraints} }\textbf{\bibinfo{volume}{6}}
 \bibinfo{pages}{345-372}

\bibitem{Smith}
\bibinfo{author}{\bibfnamefont{Smith}~\bibnamefont{B}} \bibnamefont{and}
  \bibinfo{author}{\bibfnamefont{Dyer}~\bibnamefont{M}}
  \bibinfo{year}{1996}
 \emph{  \bibinfo{journal}{Artif. Intell.}} \textbf{\bibinfo{volume}{81}}
   \bibinfo{pages}{155-181}

\bibitem{ach97}
  \bibinfo{author}{\bibfnamefont{Achlioptas}~\bibnamefont{D}},
  \bibinfo{author}{\bibfnamefont{Kirousis}~\bibnamefont{{L~M}}},
\bibinfo{author}{\bibfnamefont{Kranakis}~\bibnamefont{{E}}},
\bibinfo{author}{\bibfnamefont{Molloy}~\bibnamefont{{M}}},
  \bibnamefont{and}
  \bibinfo{author}{\bibfnamefont{Stamatiou}~\bibnamefont{Y~C}}
  \bibinfo{year}{1997}
in  \emph{\bibinfo{booktitle}{Proc. Principles and Practice of Constraint Programming}}
  \bibinfo{pages}{107-120}.

\bibitem{Gao07}
\bibinfo{author}{\bibfnamefont{Gao}~\bibnamefont{Y}} \bibnamefont{and}
  \bibinfo{author}{\bibfnamefont{Culberson}~\bibnamefont{J}}
  \bibinfo{year}{2007}
 \emph{  \bibinfo{journal}{J. Artif. Intell. Res.} }\textbf{\bibinfo{volume}{28}}
  \bibinfo{pages}{517-557}

\bibitem{smith2001}
  \bibinfo{author}{\bibfnamefont{Smith}~\bibnamefont{B}}
  \bibinfo{year}{2001}
 \emph{  \bibinfo{journal}{Theoretical Computer Science}} \textbf{\bibinfo{volume}{265}}
  \bibinfo{pages}{265-283}


\bibitem{frize}
\bibinfo{author}{\bibfnamefont{Frieze}~\bibnamefont{A}},
 \bibnamefont{and}
 \bibinfo{author}{\bibfnamefont{Molloy}~\bibnamefont{M}}
 \bibinfo{year}{2006}
  \emph{ \bibinfo{journal}{Rand. Struct. Algorithms} }\textbf{\bibinfo{volume}{28}}
  \bibinfo{pages}{323-339}


\bibitem{xu2000}
\bibinfo{author}{\bibfnamefont{Xu}~\bibnamefont{K}} \bibnamefont{and}
  \bibinfo{author}{\bibfnamefont{Li}~\bibnamefont{W}}
  \bibinfo{year}{2000}
 \emph{  \bibinfo{journal}{J. Artif. Intell. Res.}} \textbf{\bibinfo{volume}{12}}
  \bibinfo{pages}{93-103}

\bibitem{fan2011}
  \bibinfo{author}{\bibfnamefont{Y.}~\bibnamefont{Fan}},
  \bibnamefont{and}
  \bibinfo{author}{\bibfnamefont{J.}~\bibnamefont{Shen}}
  \bibinfo{year}{2011}
 \emph{  \bibinfo{journal}{Artif Intell}} \textbf{\bibinfo{volume}{175}}
  \bibinfo{pages}{914-927}

\bibitem{fan2012}
  \bibinfo{author}{\bibfnamefont{Y.}~\bibnamefont{Fan}},
  \bibinfo{author}{\bibfnamefont{J.}~\bibnamefont{Shen}},
  \bibnamefont{and}
  \bibinfo{author}{\bibfnamefont{K.}~\bibnamefont{Xu}}
  \bibinfo{year}{2012}
 \emph{  \bibinfo{journal}{Artif Intell}} \textbf{\bibinfo{volume}{193}}
  \bibinfo{pages}{1-17}

\bibitem[{\citenamefont{Zhao, Zhang, Zheng, and Xu}}]{zhao2012}
\bibinfo{author}{\bibfnamefont{C.~Y.} \bibnamefont{Zhao}},
  \bibinfo{author}{\bibfnamefont{P.}~\bibnamefont{Zhang}},
  \bibinfo{author}{\bibfnamefont{Z.~M.}~\bibnamefont{Zheng}},  \bibnamefont{and}
  \bibinfo{author}{\bibfnamefont{K.}~\bibnamefont{Xu}}
  \bibinfo{year}{2012}
 \emph{  \bibinfo{journal}{Phys. Rev. E}}\textbf{\bibinfo{volume}{85}}
  \bibinfo{pages}{016106}

\bibitem{xuwei}
\bibinfo{author}{\bibfnamefont{Xu} \bibnamefont{W}},
  \bibinfo{author}{\bibfnamefont{Zhang}~\bibnamefont{P}},
  \bibinfo{author}{\bibfnamefont{Liu} \bibnamefont{T}},\bibnamefont{and}
  \bibinfo{author}{\bibfnamefont{Gong}~\bibnamefont{F~Z}}
  \bibinfo{year}{2015}
 \emph{  \bibinfo{journal}{J. Stat. Mech.: Theory Exp.}}
   \bibinfo{pages}{P12006}



\bibitem{MMZ2005}
\bibinfo{author}{\bibfnamefont{M{\'{e}}zard}}~\bibnamefont{M},
 \bibinfo{author}{\bibfnamefont{Mora}~\bibnamefont{T}}, \bibnamefont{and}
  \bibinfo{author}{\bibfnamefont{Zecchina}~\bibnamefont{R}}
  \bibinfo{year}{2005}
  \emph{ \bibinfo{journal}{Phys. Rev. Lett.}} \textbf{\bibinfo{volume}{94}}
  \bibinfo{pages}{197205}




\bibitem{Achlioptas1}
\bibinfo{author}{\bibfnamefont{Achlioptas}~\bibnamefont{D}},
 \bibnamefont{and}
  \bibinfo{author}{\bibfnamefont{Ricci-Tersenghi}~\bibnamefont{F}}
  \bibinfo{year}{2006}
in  \emph{\bibinfo{booktitle}{Proc. STOC'06} }  pp
  \bibinfo{pages}{130-139}

\bibitem{Achlioptas2}
\bibinfo{author}{\bibfnamefont{Achlioptas}~\bibnamefont{D}},
 \bibinfo{author}{\bibfnamefont{Coja-Oghlan}~\bibnamefont{A}},
 \bibnamefont{and}
  \bibinfo{author}{\bibfnamefont{Ricci-Tersenghi}~\bibnamefont{F}}
  \bibinfo{year}{2011}
  \emph{ \bibinfo{journal}{Rand. Struct. Algorithms} }\textbf{\bibinfo{volume}{38}}
  \bibinfo{pages}{251-268}

\bibitem{Achlioptas3}
\bibinfo{author}{\bibfnamefont{Achlioptas}~\bibnamefont{D}}
\bibinfo{year}{2008}
 \emph{  \bibinfo{journal}{Eur. Phys. J. B}} \textbf{\bibinfo{volume}{64}}
  \bibinfo{pages}{395-402}

\bibitem{Achlioptas4}
  \bibinfo{author}{\bibfnamefont{Achlioptas}~\bibnamefont{D}},
    \bibnamefont{and}
  \bibinfo{author}{\bibfnamefont{Cojaoghlan}~\bibnamefont{{A}}}
  \bibinfo{year}{2010}
 in \emph{\bibinfo{booktitle}{Proc. Graph-theoretic Concepts in Computer Science}}

\end{thebibliography}

\end{document}